\documentclass{mn2e}

\newcommand{\Msolar}{\mbox{\,$\rm M_{\odot}$}}        

\usepackage{graphicx}
\usepackage{color}

\begin{document}
\title[]{The relation between the optical spectral slope and the luminosity for 17 Palomar-Green QSOs}

\author[X. Pu, W. Bian \& K. Huang]
{X. Pu$^{1}$, W. Bian$^{1,2}$, K. Huang$^{1}$ \\
$^{1}$Department of Physics and Institute of Theoretical Physics,
Nanjing Normal University, Nanjing
210097, China\\
$^{2}$Key Laboratory for Particle Astrophysics, Institute of High
Energy Physics, Chinese Academy of Sciences, Beijing 100039,
China\\
}

\maketitle

\begin{abstract}
Using 7.5-year spectroscopic monitoring data of a sample of 17
Palomar-Green QSOs (PG QSOs) (z=0.061-0.371), we obtain the optical
spectral slope for each object at all epochs by a power-law fit to
the spectra in continuum bands. All of these 17 PG QSOs exhibit
obvious spectral slope variability. Most of the 17 objects show
anti-correlation between the spectral slope and the rest-frame
5100$\AA$ continuum flux while five of them exist strong
anti-correlation (correlation coefficient R larger than 0.5). For
the ensemble of these 17 PG QSOs, a strong anti-correlation between
the average spectral slope and the average rest-frame 5100$\AA$
luminosity is found while no correlation is found between the
spectral slope and the Eddington ratio. A median anti-correlation
between spectral slope changes and continuum flux variations is also
found which indicates a hardening of the spectrum during bright
phases. Accretion disk (jet) instability models with other
mechanisms associated with changes in the accretion processes are
promising.
\end{abstract}
\begin{keywords}
galaxies:active --- galaxies:nuclei--- techniques: spectroscopic
\end{keywords}

\section{Introduction}

Variability phenomenon in active galactic nuclei (AGNs) provides a
powerful constrain on their central engine. Various classes of AGNs
show variability on timescales from hours to decades and from X-ray
to radio. The X-ray variability in AGNs provides the strong
constrain on the X-ray emission mechanics, the nucleus-size, the
supermassive black hole, et al. (e.g. Mushotzky et al. 1993; Bian \&
Zhao 2003). The large-amplitude and short-timescale optical
variability found in Blazars are likely due to the relativistic beam
effects (e.g. Fan \& Lin 2000; Vagnetti et al. 2003). For the
small-amplitude and long-timescale optical variability in AGNs, its
origin is still a question to debate and there are mainly four
models: accretion disk instabilities (e.g. Rees 1984; kawaguchi et
al. 1998), supernovae explosions (e.g. Cid Fernandes et al. 1996;
Aretxaga et al. 1997), gravitational lensing (e.g. Hawkins 1996),
and star collisions (e.g. Torricelli-Ciamponi et al. 2000). It is
still difficult to disentangle various models.

For a long time the long-term multi-wavelength photometric
monitoring plays an important role in obtaining the optical
variability of QSOs (see Table 1 in Giveon et al. 1999). Many
results on correlations between the variability and the luminosity,
the redshift, the time lag, the rest-frame wavelength, spectral
properties are obtained (e.g. Huang et al. 1990; Giallongo et al.
1991, Cimatti et al. 1993, Cristiani et al. 1996, Giveon et al.
1999, Trevese et al. 2002, Vanden Berk et al. 2004). Using the
photometric data in various optical bands, the spectral variability
can also be studied. Giveon et al.(1999) found that about half of
the 42 Palomar-Green QSOs (PG QSOs) become bluer (harder) when they
brighten (see also Trevese et al. 2001). Recently, using the
photometric data from the Sloan Digital Sky Survey (SDSS), Vanden
Berk et al.(2004) found quasars are systematically bluer when
brighter at all redshifts for a sample of over 25000 QSOs (see also
de Vries et al. 2003).

Although the photometric method have some advantages that the
variability amplitude is accurate and that it takes less time
comparing with spectroscopic method, it just monitors the flux
variability in a few points in the wavelength which usually consists
of many components including continuum and strong emission lines. So
it is necessary to study the optical spectral variability using
spectroscopic monitoring data. Recently, Wilhite et al.(2005)
presented 315 significantly varied quasars sample from multi-epoch
spectroscopic observations of SDSS. They found that the average
difference spectrum (bright phase minus faint phase) is bluer than
the average single-epoch quasar spectrum. However, the epochs are
just a few ($\sim2$) for these 315 SDSS QSOs.

The AGNs spectral variability has focused on about ~30 individual
objects for a long-term of decades, with the primary goals to
obtain the structure of the broad line regions (BLRs) and the
masses of their central supermassive black holes (Reverberation
mapping methods, Peterson 1993). Kaspi et al. (2000) presented
results from a spectrophotometrically monitoring program of a
well-defined sample of 28 PG QSOs. 17 of these 28 PG QSOs were
observed with good time sampling ($\geq 20$ observing epochs).
Good time sampling and long-term monitoring make these 17 QSOs the
best objects in the studying of the spectral slope variability. In
section 2, the sample, observations and data reduction are briefly
described. The spectroscopic data analysis is given in section 3.
Our discussions and results are presented in the last section. All
of the cosmological calculations in this paper assume $H_0= 70
{\rm ~km~s^{-1}~Mpc^{-1},~q_0=0.5}$.

\section{OBSERVATIONS AND DATA REDUCTION}
\label{sect:Obs}

The 7.5-year spectra of 17 PG QSOs are available on the web
{\footnote {http://wise-obs.tau.ac.il/\~{}shai/PG/}}. The sample,
observing technique, and reduction procedure are briefly described
here. The optical spectrophotometric observations of 17 PG QSOs
were done using 2.3 m telescope at Steward Observatory (SO) and 1
m telescope at Wise Observatory (WO). The observations were
performed between 1991 and 1998. The typical spectrum wavelength
coverage at both observatories is from 4000 to 8000$\AA$ with a
spectral resolution of about 10$\AA$. The redshifts of all these
17 QSOs are less than 0.4 and $B < 16$ mag. Total exposure times
were usually 40 minutes at SO 2.3 m telescope and 2 hours at WO 1
m telescope. Spectrophotometry was obtained every 1-4 months for
7.5-year per object. Standard IRAF routines were used to perform
spectroscopic data reduction. The consecutive quasar/star flux
ratios were compared to test for systematic errors in the
observations. Spectrophotometric calibration for each quasar was
accomplished excellently and the accuracies of order 1-2\% can be
achieved. Spectra were calibrated to an absolute flux scale using
observations of spectrophotometric standard stars on one or more
epochs. These 17 PG QSOs are primarily used to obtain BLRs size
and the masses of their central supermassive black holes (Kaspi et
al. 2000). The observed spectra have been corrected for Galactic
extinction using $A_{B}$ values from NED (See table 1 in Kaspi et
al. 2000) {\footnote {The NASA/IPAC Extragalactic Database (NED)
is operated by the Jet Propulsion Laboratory, California Institute
of Technology, under contract with the National Aeronautics and
Space Administration.}}, assuming an extinction curve with $R_V =
3.1$. Most of these 17 QSOs are radio-quiet, and there are two
radio-loud QSOs: PG 1226+023 (3C 273), PG 1704+608 with the radio
loudness of 1621, 563, respectively (Nelson 2000). In Table 1, the
object name, redshift, apparent B magnitude, number of
spectrophotometric observing epochs are listed.

\section{DATA ANALYSIS}
\label{sect:Ana}

Our goal is to investigate the long-term spectral variability of the
PG sample. It is popularly accepted that we can use the power law
formulae, $f_\nu \propto \nu ^{-\alpha} $ ($f_\lambda \propto
\lambda ^{-2+\alpha} $), to approximately fit the optical continuum
spectrum of AGNs (e.g. Wilhite et al.2005). Some authors also used
two power laws to model the continuum emission(e.g. Forster et al.
2001; Shang et al. 2005). The spectral slope usually changes with
luminosity, which can naturally explain the relation between the
variability and the wavelength. From the spectra of these 17 PG QSOs
during 7.5-year spectroscopic observations, we can obtain the
spectral slope $\alpha$ ($f_\nu \propto \nu ^{-\alpha} $) by fitting
a power law to the continuum spectrum, using spectral regions
unaffected by other emission components. The usually used "continuum
windows" (at the rest-frame) known to be relatively free from strong
emission lines are 3010-3040, 3240-3270, 3790-3810, 4200-4230,
5080-5100, 5600-5630, 5970-6000, and 6005-6035\AA (Forster et al.
2001, Vanden Berk et al. 2001). Here we directly used the continuum
bands in the observer's frame defined by Kaspi et al. (2000) (See
table 2 in Kaspi et al. 2000). The Balmer lines and other strong
broad emission lines are excluded in these continuum bands. Most of
these 17 PG QSOs have 6 continuum bands. Two objects, PG 1613+658
and PG 1700+518, have three continuum bands. From the best fit, we
derive the spectral slope $\alpha$, the continuum flux at rest-frame
5100$\AA$ and their errors for all observing epochs per object. The
average spectral slope and average flux density at rest-wavelength
5100$\AA$ for each of these 17 PG QSOs are listed in Col. (5) and
(6) in table 1. Here we mainly study the relation between $\alpha$
and continuum flux $\log f_{5100}$.

In Fig. 1, we showed the spectral slope $\alpha$ as a function of
the rest-frame 5100$\AA$ continuum flux $\log f_{5100}$ for each
object. Considering the errors, it is obvious that all of these 17
PG QSOs exhibit spectral slope variability. We used the simple
least-squares linear regression (Press et al. 1992) to study the
correlation per object. For each object, we obtained the correlation
coefficients (R) and the probabilities (P) for rejecting the null
hypothesis of no correlation. The fitting results for all spectral
data per object are listed in Col. (7) and (8) in table 1. These
best fittings are also plotted in each panel (Fig. 1, solid lines).
From Fig. 1 and table 1, we found that most of the objects showed
negative correlation coefficients. Among the 15 QSOs which showed
negative correlations, five objects PG 0026+129, PG 0052+251, PG
0804+761, PG 1617+175, and PG 1704+608 showed strong
anti-correlations with $R > 0.5$. Two objects, PG 1229+204 and PG
1700+518, existed moderate opposite trends. One radio-loud QSO, PG
1704+608, showed the strongest anti-correlation with $R = -0.88$.
While the other radio-loud QSO, PG 1226+023 (3C 273), showed weaker
anti-correlation with $R = -0.23$.

In order to give the ensemble relation for these 17 PG QSOs between
the spectral slope and the luminosity, in the left panel in Fig. 2,
we showed the average spectral slope $\alpha$ as a function of the
average rest-frame 5100$\AA$ luminosity $\log L_ {5100}$. We found
that there existed a strong anti-correlation (R=-0.57, P=0.016). A
simple least-squares linear regression gives the relation,
\begin{equation}
\alpha=(-0.38\pm0.07)\log L_{5100}+(11.97\pm2.08)
\end{equation}
The best fitting is showed as solid line in Fig. 2. We also
calculated the average Eddington ratio for each object,
$L_{bol}/L_{Edd}$. The bolometric luminosity $L_{bol}$ is from
luminosity in rest-frame 5100$\AA$: $L_{bol}=9\lambda L_{\lambda}
(5100\AA)$, and $\rm L_{Edd}=1.26\times 10^{38}M_{BH}/\Msolar~ergs~
s^{-1}$, where the black hole mass is from Kaspi et al. (2000). In
the right panel of Fig. 2, we plotted average $\alpha$ versus
average Eddington ratio for each object. We found no correlation
between them (R=0.064, P=0.081).

Using the fitting slope of these 17 PG QSOs at all epochs, we
correlated the spectral slope changes with the continuum flux
variations. The changes of the spectral slope are calculated for
each object between the slope of each observing date and the average
spectral slope. In Fig. 3, the overall $\delta \alpha $ versus
$\delta \log f_ {5100}$ is plotted. We found an anti-correlation
between the spectral slope changes and the continuum flux variations
with $R=-0.27$ and $P<10^{-4}$. A simple least-squares linear
regression gives,
\begin{equation}
\delta\alpha=(-0.61\pm0.01)\times\delta\log
f_{5100}-(5.91\pm0.52)\times10^{-3}
\end{equation}
The best fitting is showed as the solid line in Fig. 3.

\section{DISCUSSIONS AND RESULTS}
\label{sect:Dis}

These 17 PG QSOs data span 7.5 years with more than 20 epochs and
are used in the reverberation mapping method to determine their
central supermassive black hole masses (e.g. Peterson 1993; Kaspi et
la. 2000). The time-span is enough to determine the variability of
the optical spectral slope if any. For these high-luminosity PG
QSOs, the contribution from the host galaxies is negligible. Here we
used a power law function to fit data in the continuum bands
suggested by kaspi et al. (2000). And then we obtained the spectral
slope and the flux $f_{5100}$ in the rest-frame. Their errors are
from our fitting. The results would possibly depend on the selection
of the continuum bands. In order to be consistent with the continuum
analysis of Kapsi et al. (2000), We adopted continuum bands
suggested by kaspi et al. (2000) instead of usual continuum
windows¡±(Forster et al. 2001, Vanden Berk et al. 2001). In the
future we will use the multi-component fitting method to avoid the
selection effect of the ¡±continuum windows¡± (Bian, Yuan \& Zhao
2005).

\subsection{Correlation for individual objects}
\label{individual}

Giallongo et al.(1991) investigated the long-term optical
variability of quasars from photometric observations and found a
positive correlation between variability and redshift. That QSOs
at higher redshifts show larger variability since they are
observed at a higher rest-frame frequency where the variability is
stronger was indicated. This suggests a hardening in the bright
phase of quasars. Statistical evidence for this trend were
presented by follow-up studies (e.g. Cristiani et al. 1996, Di
Clemente et al. 1996). Giveon et al.(1999) presented results from
7 yr photometric monitoring program of a well-defined, optically
selected sample of 42 PG QSOs. The spectra of about half of the
QSOs in their sample became harder when they brighten. The 42
objects in the optical sample included all 17 objects discussed in
the present paper. Here on the basis of 7.5-year spectroscopic
monitoring data of these 17 PG QSOs, we correlated the spectral
slope with the rest-frame 5100$\AA$ continuum flux. The results
showed that, among these 17 PG QSOs, five objects displayed strong
anti-correlation with $R > 0.5$ and  there are nine objects with
$R > 0.3$ (See Fig. 1 and table 1). The anti-correlation implied
that QSOs become bluer as they brighten. At the same time, other
objects showed weaker correlation between $\alpha$ and $f_{5100}$
(See Fig. 1 and table 1). For PG 1700+518 and PG 1229+204, the
correlation is possibly positive. The Fe II multiples are strong
in PG 1700+518 (see Fig. 1 in Kaspi 2000). The strong Fe II
emission may contaminate the selected continuum bands and the
studying of the spectral slope variability for this object is
probably influenced. The spectra of these 17 PG QSOs are from 2.3
m telescope at Steward Observatory and 1 m telescope at Wise
Observatory. For each object, a large fraction of the spectral
data are from the Wise Observatory. Considering only the spectra
from the WO, a strong anti-correlation between the spectral slope
and the continuum flux is found in PG 1229+204 ($R=-0.80$,
$P<10^{-4}$) (see Fig. 1). The least-squares linear fitting using
the spectral data only from the WO are also showed for all the
other PG QSOs in our sample (Fig. 1, dotted lines). The result for
this fitting are listed in Col. (9) and (10) in table 1. Using the
spectra only from the WO, all but the object PG 1700+518 showed
anti-correlation between the spectral slope and the continuum
flux. Stronger anti-correlation are obtained for some of the
objects (e.g. PG 1426+015, PG 1613+658, PG 2130+099).

\subsection{Correlation for the ensemble of 17 objects}
\label{ensemble}

In order to clarify the slope variability, some authors discussed
the relation between the spectral energy distribution (SED) changes
and the continuum flux variations. Using the variation between two
epochs in the photographic U, B$_ J$, F, and N bands, Trevese et
al.(2001) correlated the spectral slope changes $\Delta \alpha$ with
brightness variations $\Delta \log f_{J}$ for a complete
magnitude-limited sample of faint quasars in SA 57 and detected an
average increase of the spectral slope (note that their $\alpha$ is
defined by $f_\nu \propto \nu ^{\alpha} $) for increasing continuum
flux, indicating hardening of the spectrum in the bright phases.
Trevese et al.(2002) performed an analysis of B and R observing data
of 42 PG quasars from Giveon et al.(1999). They showed in their
results that the average spectral slope of each QSO tended to be
larger for brighter objects (the $\alpha$ is defined by $f_\nu
\propto \nu ^{\alpha} $). Using 7.5-year spectroscopic observing
data, we discussed the relation between $\Delta \alpha$ and $\Delta
\log f_{5100}$ for the ensemble of these 17 PG QSOs, a median
anti-correlation ($R=-0.27$, $P<10^{-4}$) between them is found (see
Fig. 3), indicating a hardening of the spectrum during bright phases
which is consistent with the result of Trevese et al.(2001). We also
found a strong anti-correlation between the average spectral slope
and the average rest-frame 5100$\AA$ luminosity (See the left panel
in Fig. 2) which is qualitatively consistent with the inter-QSO
result of Trevese et al. (2002).

\subsection{Model behind optical spectral variability}
\label{model}

The physical mechanisms behind the optical variability of AGNs is
largely unknown. Some models have been proposed to explain the AGNs
optical variability properties. Current models can be classified
mainly into three groups: accretion disk instabilities,
discrete-event or Poissonian processes, and gravitational
microlensing (e.g. Hawkins 1996, Aretxaga et al. 1997, kawaguchi et
al. 1998, Kong et al. 2004). The optical spectral variability can
provide clues to disentangle these models. Using the optical spectra
of NGC 5548 between 1989 and 2001, Kong et al.(2004) found a strong
correlation between $\alpha$ and the luminosity at rest-frame
5100$\AA$. They used the global variance of the accretion rates
or/and the variance of the inner radius of the accretion disk to
account for the strong relation. It can explain well the
anti-correlation between the optical spectral index and the
continuum luminosity observed in NGC 5548. Here in our PG QSOs
sample, this disk instability model can also explain the
anti-correlation which most of the objects showed between the
spectral slope $\alpha$ and the rest-frame 5100$\AA$ continuum flux
for individual QSOs. However, for the ensemble of our PG QSOs
sample, the global variance of the accretion rates would not
necessarily account for the global slope variance (See the right
panel in Fig. 2). Vanden Beck et al. (2004) studied the ensemble
photometric variability for about 25000 SDSS QSOs, that quasars are
systematically bluer when brighter at all redshifts were found.
Their result also seemed to disfavor gravitational microlensing and
generic Poissonian processes as the primary source of quasar
variability. QSOs are widely believed to be powered by the accretion
disk onto a supermassive black hole (e.g. Rees 1984). It is natural
to consider that the QSOs variability is due to some mechanisms
associated with changes in the accretion processes. Kawaguchi et al.
(1998) presented a very simple cellular-automaton model for disk
instability. They gave the slope of structure function between 0.41
to 0.49, which is inconsistent with SDSS results (Vanden Beck et al.
2004). It is likely due to the complexity of possible accretion disk
(or jet) instability models, which prevented more quantitative
predictions.

\section{Conclusion}

Using 7.5-year spectroscopic monitoring data of a sample of 17 PG
QSOs, we study the optical spectral slope variance. The main
conclusions can be summarized as follows:

\begin{itemize}

\item{Using the continuum bands suggested by Kaspi et al.
(2000), we found that, in 7.5-years long-term observation, all 17 PG
QSOs showed obvious optical spectral slope variability.}

\item{Most of these 17 PG QSOs showed anti-correlation between the
spectral slope $\alpha$ and the rest-frame 5100$\AA$ continuum flux
$\log f_{5100}$ while five of them showed strong anti-correlation
between $\alpha$ and $\log f_{5100}$ ($R > 0.5$).}

\item{For the ensemble of these 17 PG QSOs, a strong
anti-correlation ($R = -0.57$) between the average spectral slope
$\alpha$ and the average rest-frame 5100$\AA$ luminosity $\log L_
{5100}$ is found while a median anti-correlation ($R = -0.27$) is
found between spectral slope changes $\delta \alpha $ and continuum
flux variations $\delta \log f_ {5100}$ indicating a hardening of
the spectrum during bright phases.}

\item{The disk instability model can qualitatively explain the
anti-correlation between the spectral slope $\alpha$ and the
continuum flux $\log f_{5100}$ for individual QSOs. However, the
global variance of the accretion rates would not necessarily account
for the global slope variance for the ensemble of our PG QSOs
sample. Accretion disk (jet) instability models with other
mechanisms associated with changes in the accretion processes are
promising.}

\end{itemize}

\section*{ACKNOWLEDGMENTS}
This work has been supported by the NSFC (No.10403005;
No.10473005) and the science-technology key foundation from
Education Department of P. R. China (No. 206053).

\begin{table*}
  \caption{Sample properties. Col. 1:The name, Col. 2: redshift,
  Col. 3: apparent B magnitude, Col. 4: number of spectrophotometric observing epochs, Col. 5: the average optical
  spectral slope, Col. 6: average continuum
  flux density at rest wavelength 5100$\AA$ in units of
  $10^{-16}ergs ~s^{-1}cm^{-2}\AA^{-1}$, Col. 7-8: correlation
  coefficient and probability for all spectra per object, Col. 9-10: correlation
  coefficient and probability for the spectra only from the Wise Observatory.}
  \label{}
  \begin{center}\begin{tabular}{cccccccccccccccc}
  \hline\noalign{\smallskip}
Object & z & $m_B$ & N & $\alpha$ & $f_{\lambda}$ & R & P  & $R_{Wise}$ & $P_{Wise}$  \\
(1) & (2) & (3) & (4) & (5) & (6)& (7) & (8) & (9) & (10)  \\
  \hline\noalign{\smallskip}
PG 0026+129 & 0.142 & 15.3 & 56  & $1.026\pm0.116$ & $27.4\pm4.0$ & -0.51 & $<1\times10^{-4}$ & -0.62 & $<1\times10^{-4}$ \\
PG 0052+251 & 0.155 & 14.7 & 56  & $0.978\pm0.229$ & $22.4\pm4.4$ & -0.65 & $<1\times10^{-4}$ & -0.70 & $<1\times10^{-4}$ \\
PG 0804+761 & 0.100 & 14.5 & 70  & $0.290\pm0.097$ & $55.7\pm9.5$ & -0.70 & $<1\times10^{-4}$ & -0.69 & $<1\times10^{-4}$ \\
PG 0844+349 & 0.064 & 15.1 & 49  & $1.047\pm0.157$ & $37.8\pm3.4$ & -0.22 & 0.13 & -0.38 & 0.024 \\
PG 0953+414 & 0.239 & 15.4 & 36  & $0.117\pm0.101$ & $15.7\pm2.1$ & -0.19 & 0.27 & -0.42 & 0.046 \\
PG 1211+143 & 0.085 & 14.4 & 38  & $0.590\pm0.177$ & $58.1\pm10.4$ & -0.39 & 0.017 & -0.35 & 0.050 \\
PG 1226+023 & 0.158 & 12.8 & 39  & $0.186\pm0.091$ & $218\pm26$ & -0.23 & 0.16 & -0.083 & 0.66 \\
PG 1229+204 & 0.064 & 15.5 & 33  & $1.149\pm0.189$ & $27.9\pm4.3$ & 0.38 & 0.028 & -0.80 & $<1\times10^{-4}$ \\
PG 1307+085 & 0.155 & 15.6 & 23  & $0.564\pm0.159$ & $18.5\pm2.0$ & -0.34 & 0.12 & -0.25 & 0.36 \\
PG 1351+640 & 0.087 & 14.7 & 30  & $0.637\pm0.081$ & $52.0\pm5.0$ & -0.49 & $5.7\times10^{-3}$ & -0.52 & 0.010 \\
PG 1411+442 & 0.089 & 15.0 & 24  & $0.685\pm0.080$ & $36.8\pm3.1$ & -0.10 & 0.63 & -0.29 & 0.22 \\
PG 1426+015 & 0.086 & 15.7 & 20  & $0.575\pm0.166$ & $46.3\pm6.7$ & -0.17 & 0.49 & -0.65 & $4.5\times10^{-3}$ \\
PG 1613+658 & 0.129 & 14.9 & 48  & $0.423\pm0.149$ & $34.5\pm4.4$ & $-4.9\times10^{-3}$ & 0.97 & -0.74 & $<1\times10^{-4}$ \\
PG 1617+175 & 0.114 & 15.5 & 35  & $0.524\pm0.316$ & $13.8\pm2.7$ & -0.57 & $3.4\times10^{-4}$ & -0.70 & $<1\times10^{-4}$ \\
PG 1700+518 & 0.292 & 15.4 & 39  & $0.828\pm0.160$ & $21.6\pm1.5$ & 0.27 & 0.10 & 0.20 & 0.29 \\
PG 1704+608 & 0.371 & 15.6 & 25  & $0.733\pm0.188$ & $16.9\pm2.4$ & -0.88 & $<1\times10^{-4}$ & -0.85 & $1.1\times10^{-4}$ \\
PG 2130+099 & 0.061 & 14.7 & 64  & $0.673\pm0.150$ & $50.6\pm4.7$ & -0.049 & 0.70 & -0.57 & $<1\times10^{-4}$ \\
  \noalign{\smallskip}\hline
  \end{tabular}\end{center}
\end{table*}

\newpage
\begin{figure*}
   \centering
   \includegraphics[width=80mm]{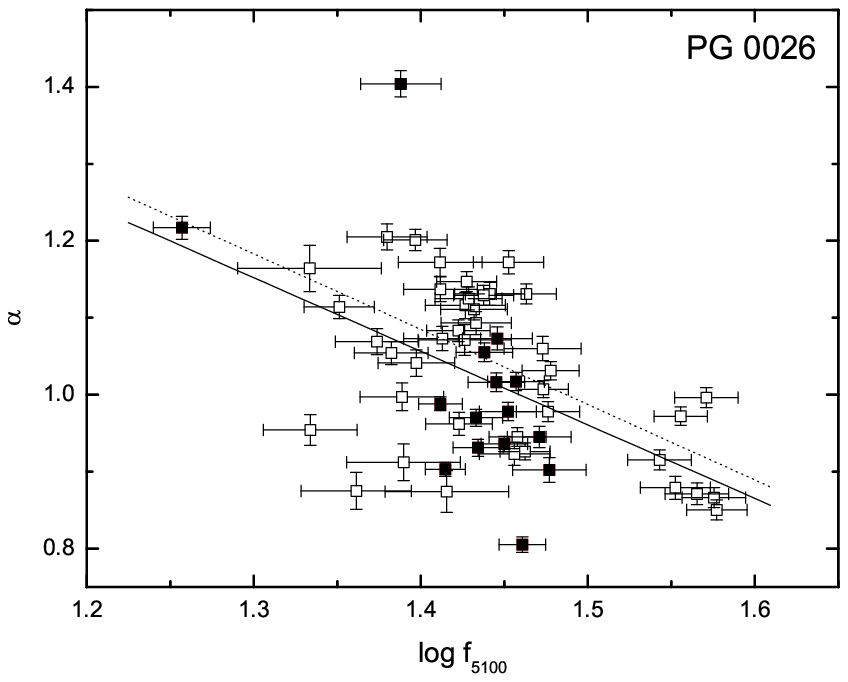}
   \includegraphics[width=80mm]{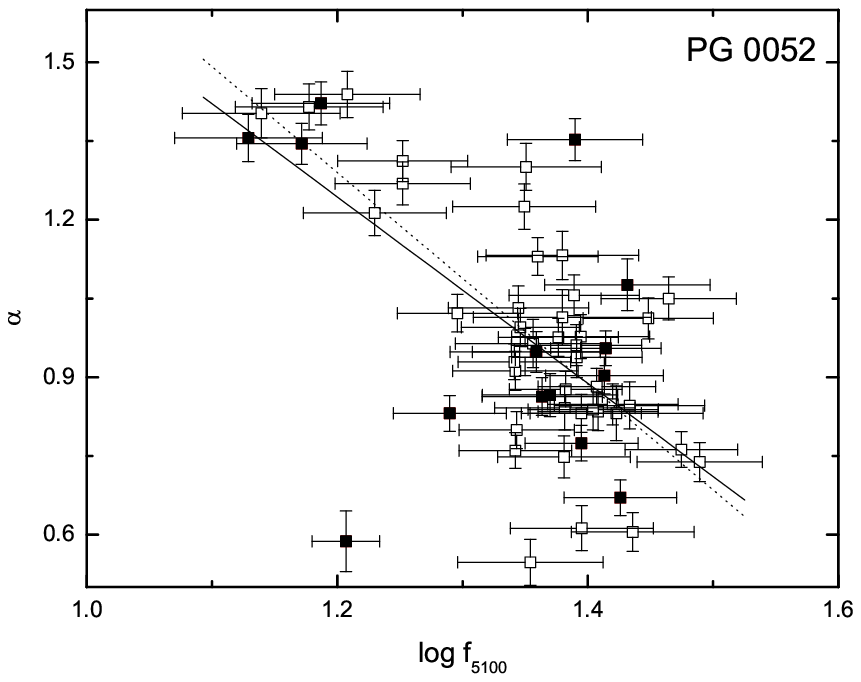}
   \includegraphics[width=80mm]{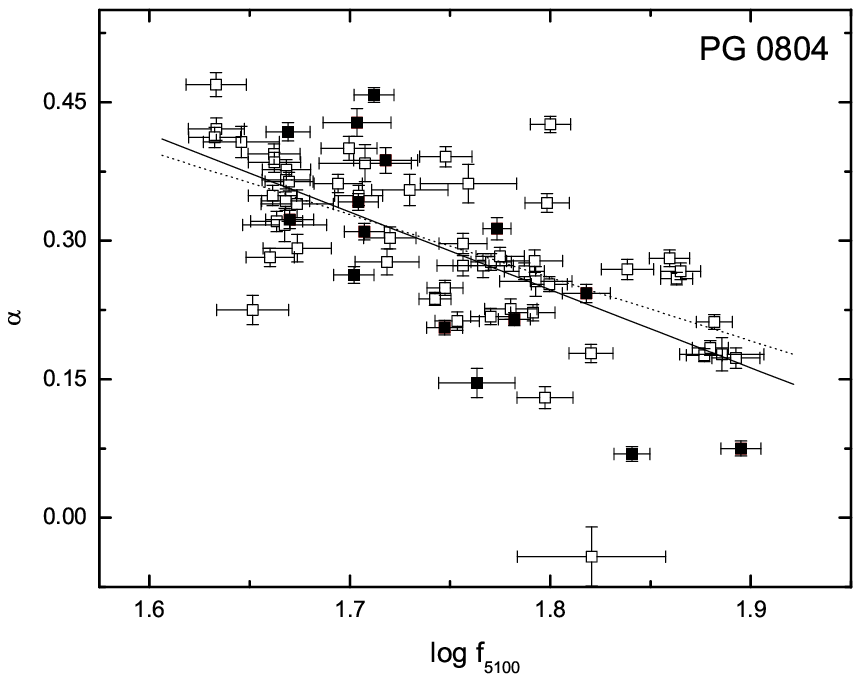}
   \includegraphics[width=80mm]{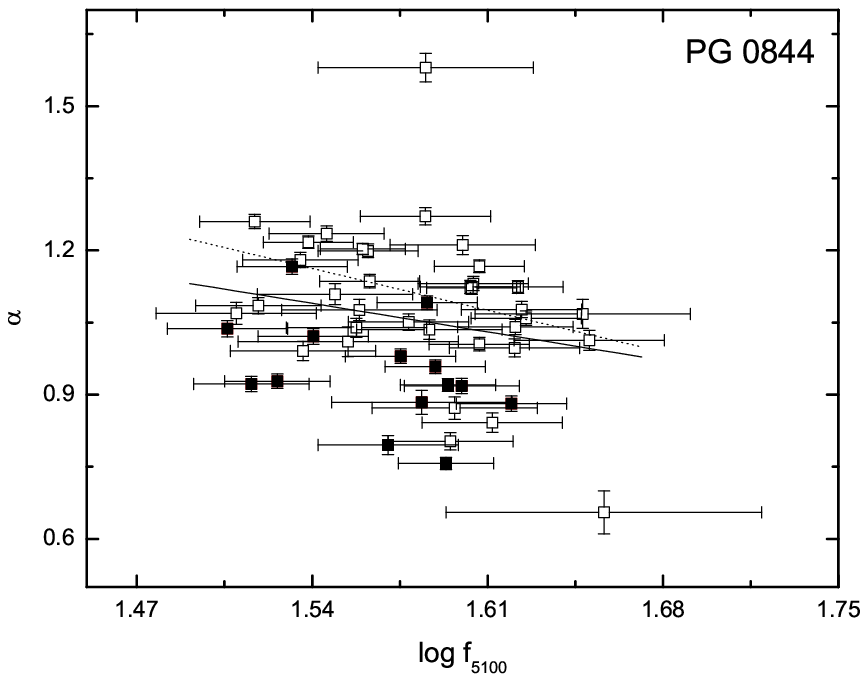}
   \includegraphics[width=80mm]{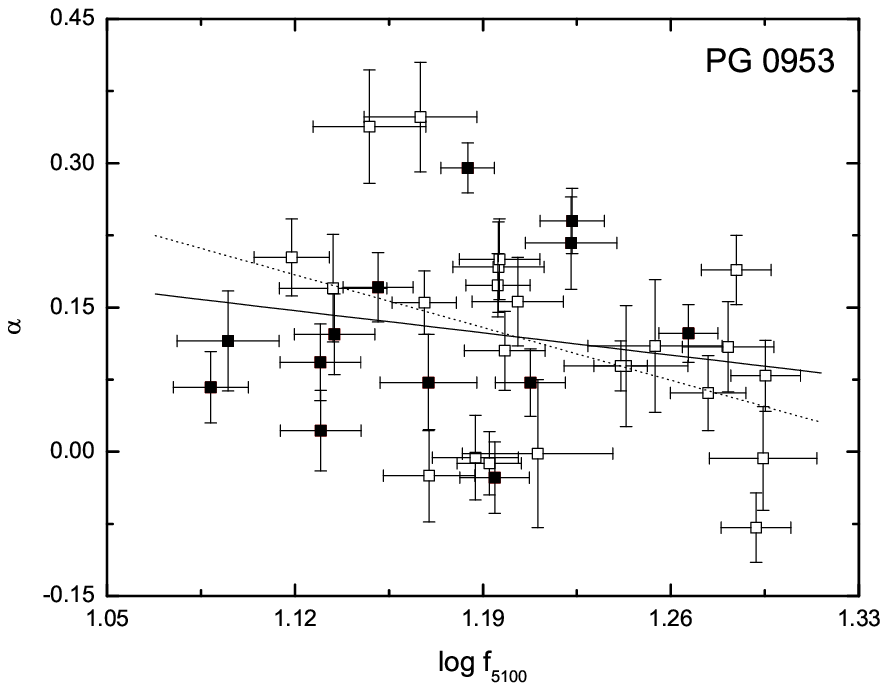}
   \includegraphics[width=80mm]{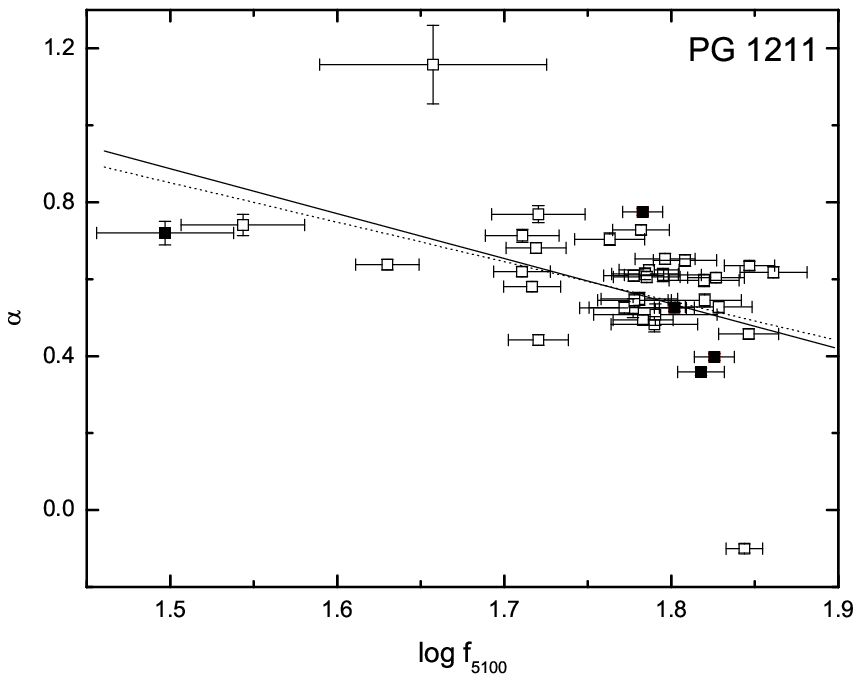}
   \caption{The spectral slope $\alpha$ versus the rest-frame 5100$\AA$
continuum flux $\log f_ {5100}$ (in units of $10^{-16}ergs
~s^{-1}cm^{-2}\AA^{-1}$) for each of these 17 PG QSOs. The
spectral data from the Steward Observatory are solid rectangles,
and the spectral data from the Wise Observatory are open
rectangles, respectively. Solid lines are the best fit to all
spectral data for each object while dotted lines are the best fit
to the spectral data only from the Wise Observatory.}
   \label{}
   \end{figure*}

\setcounter{figure}{0}
\begin{figure*}
   \centering
   \includegraphics[width=80mm]{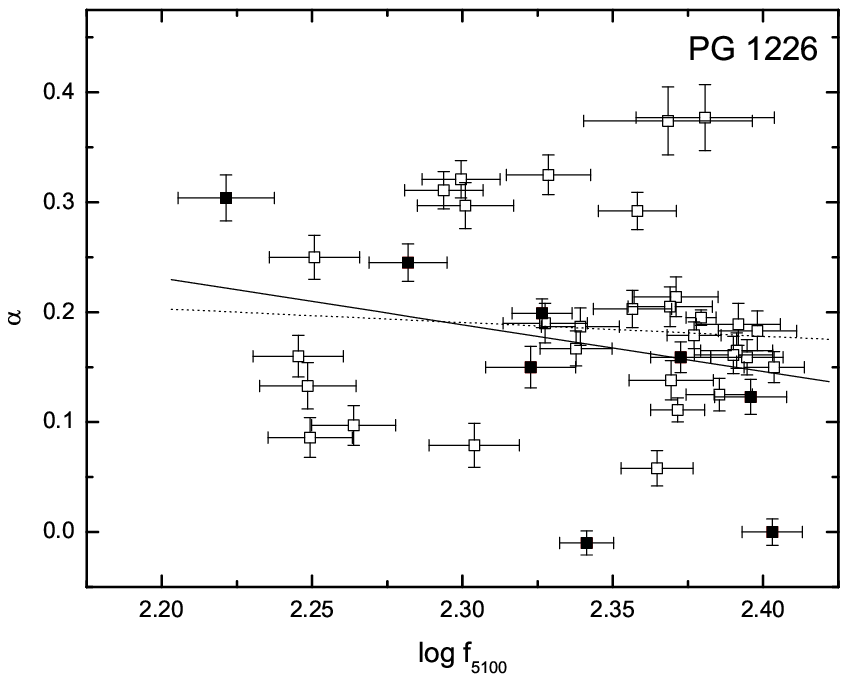}
   \includegraphics[width=80mm]{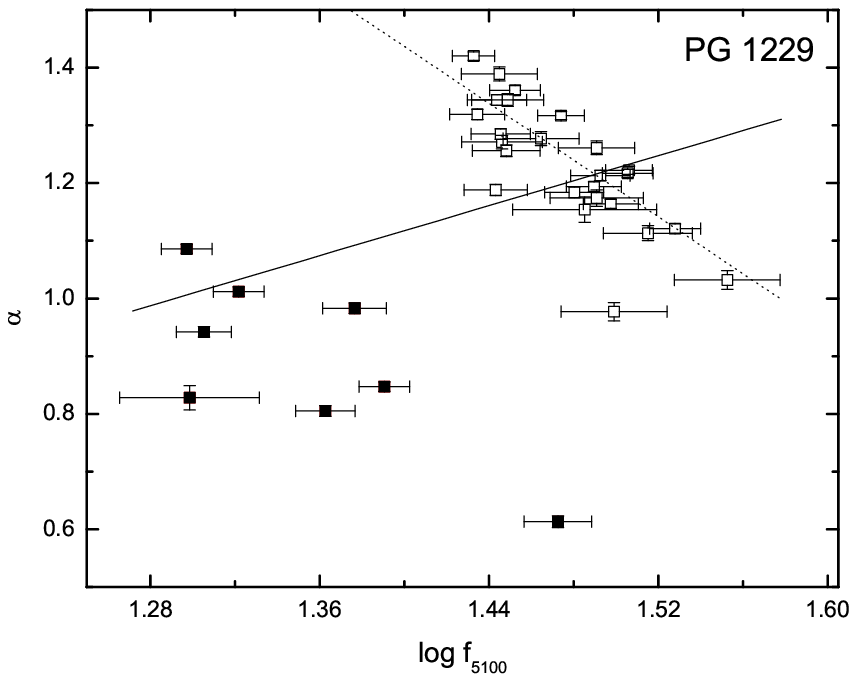}
   \includegraphics[width=80mm]{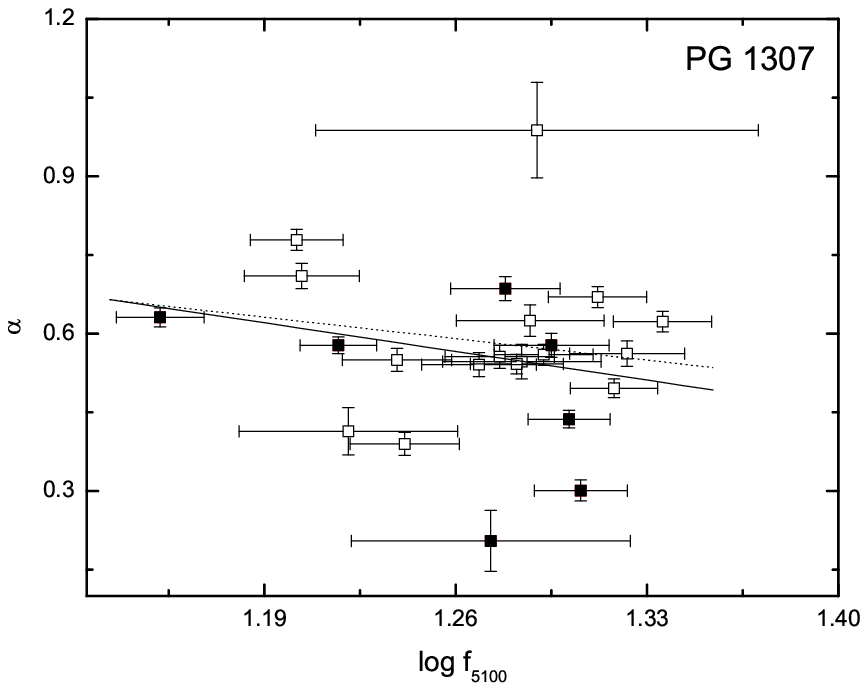}
   \includegraphics[width=80mm]{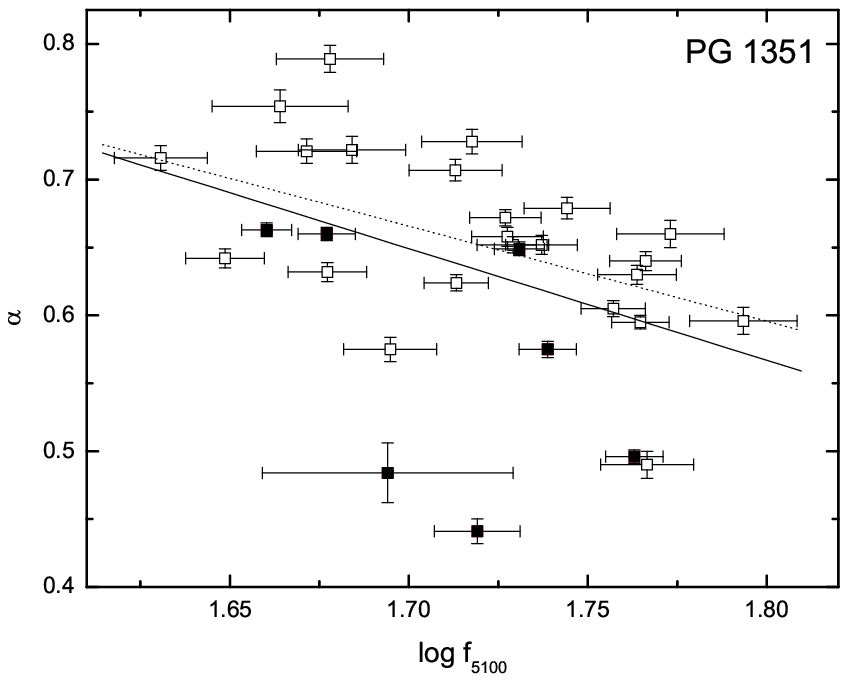}
   \includegraphics[width=80mm]{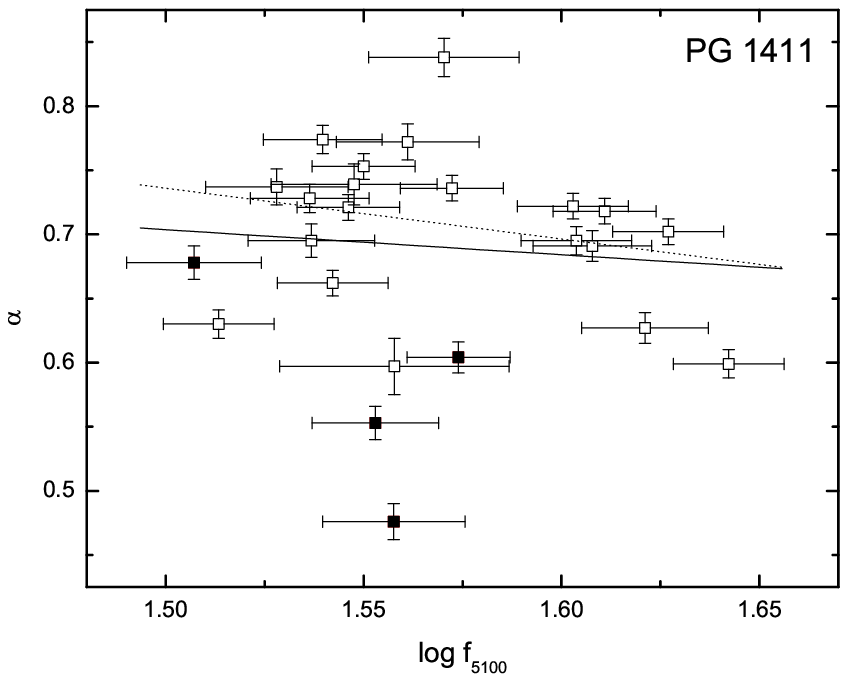}
   \includegraphics[width=80mm]{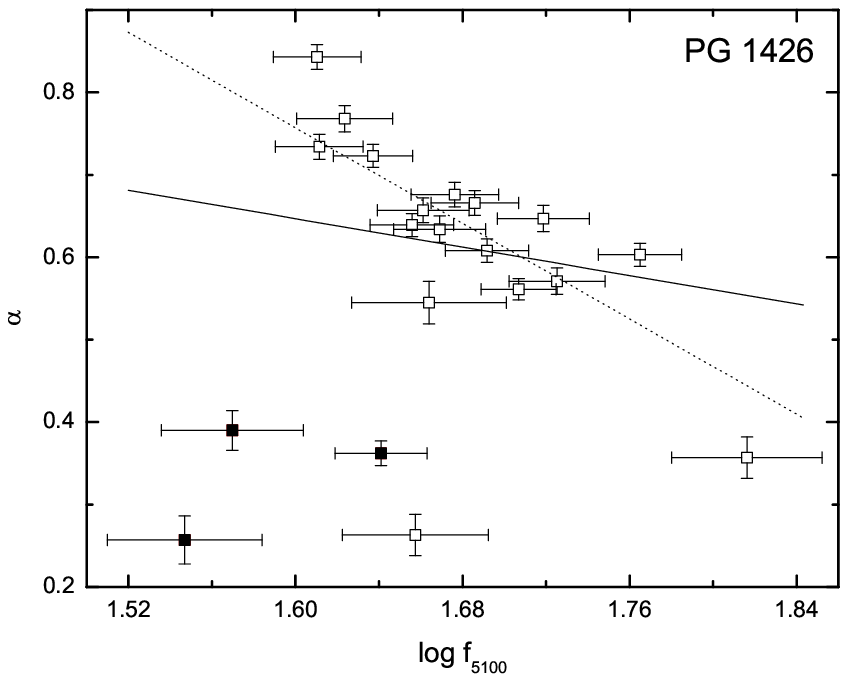}
   \caption{Continued.}
   \label{}
   \end{figure*}

\setcounter{figure}{0}
\begin{figure*}
   \centering
   \includegraphics[width=80mm]{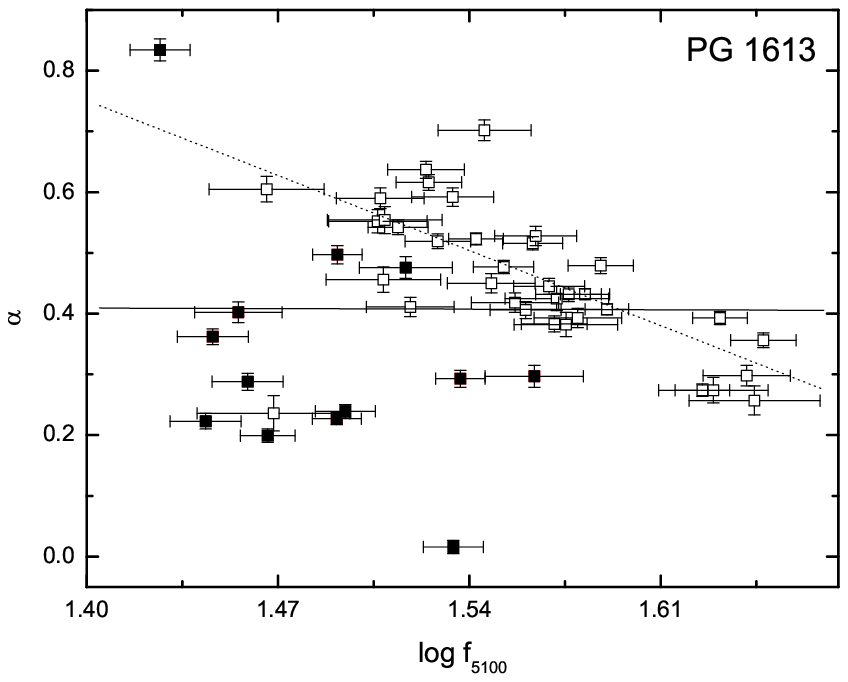}
   \includegraphics[width=80mm]{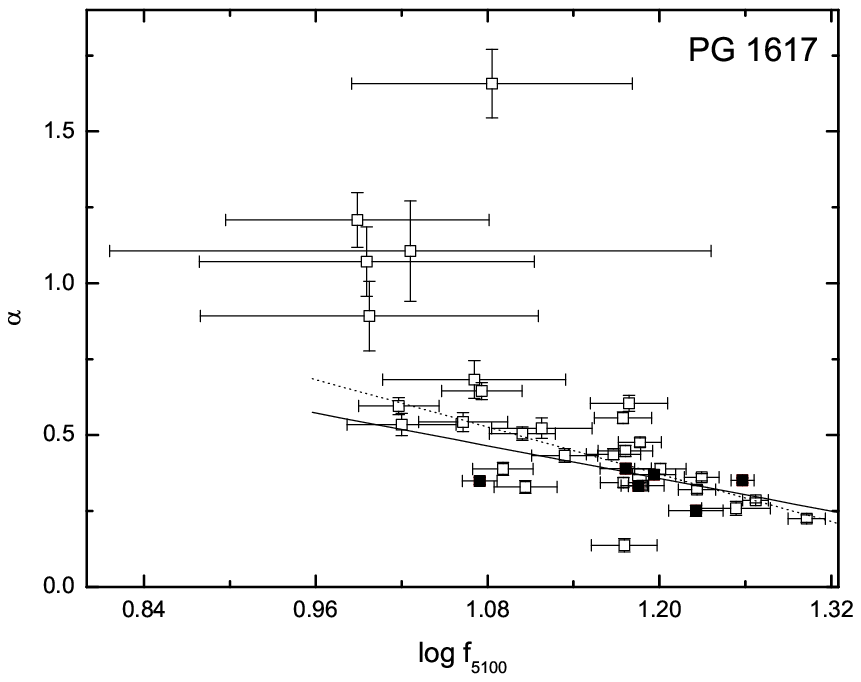}
   \includegraphics[width=80mm]{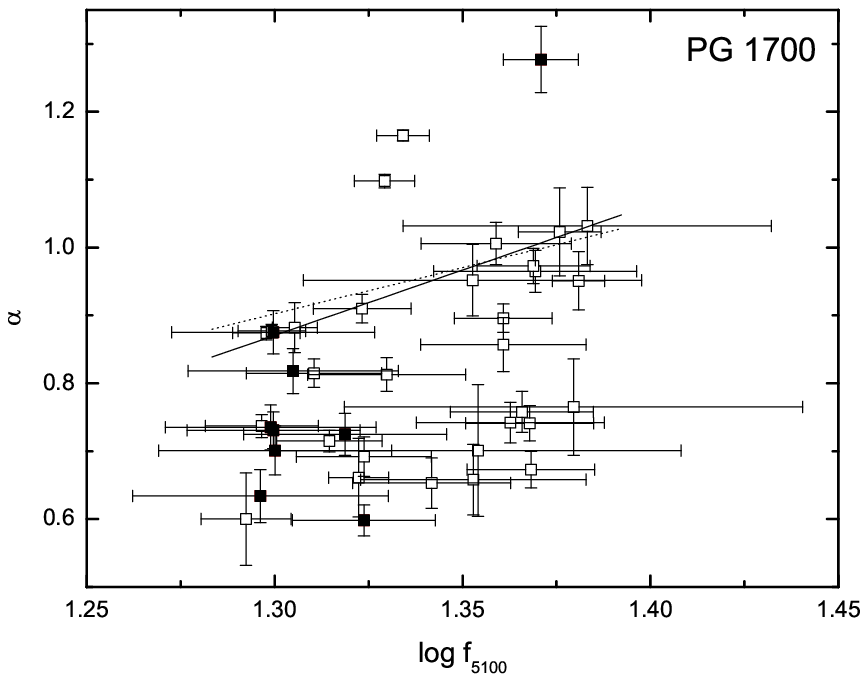}
   \includegraphics[width=80mm]{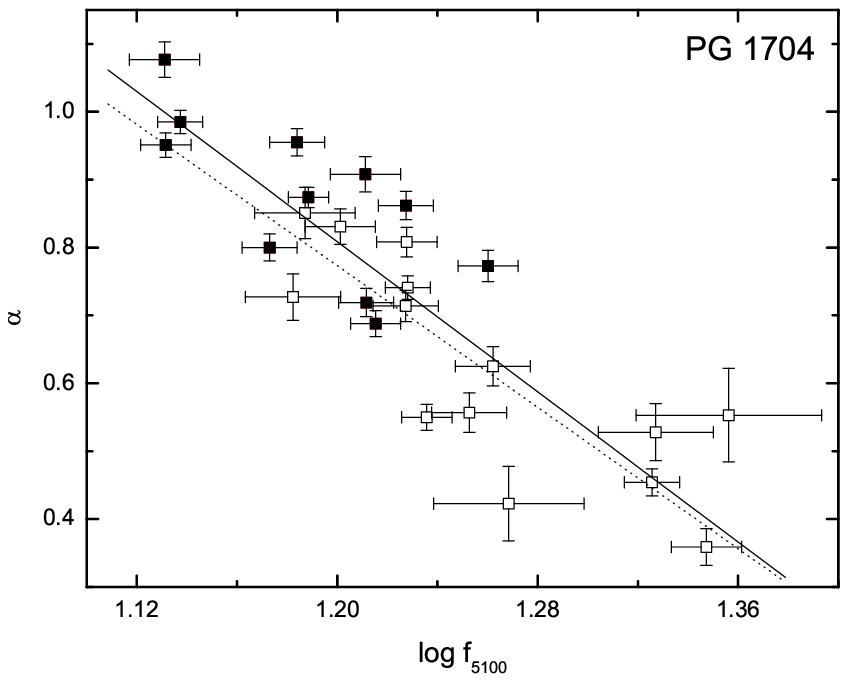}
   \includegraphics[width=80mm]{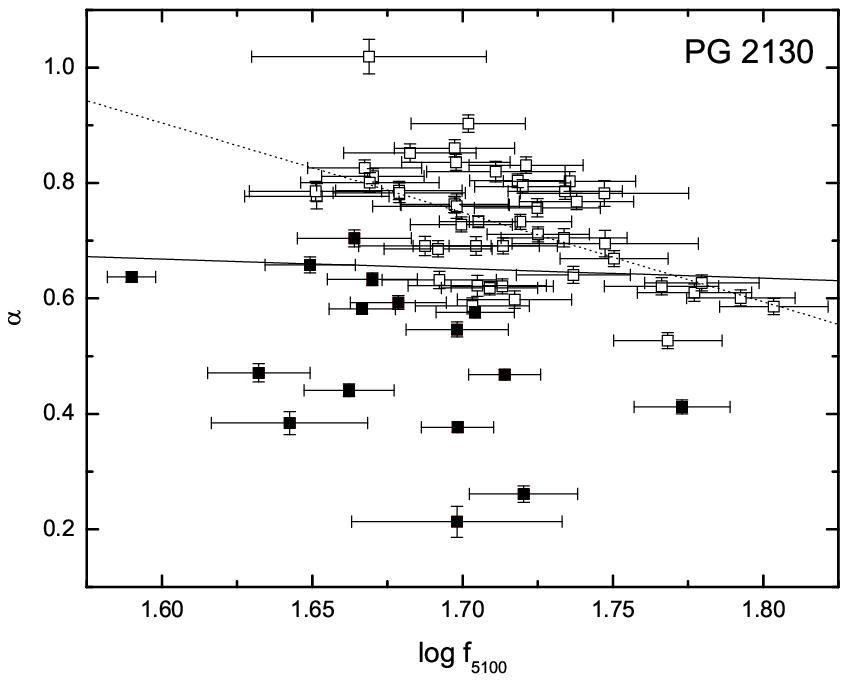}
   \caption{Continued.}
   \label{}
   \end{figure*}

\begin{figure*}
   \centering
   \includegraphics[width=80mm]{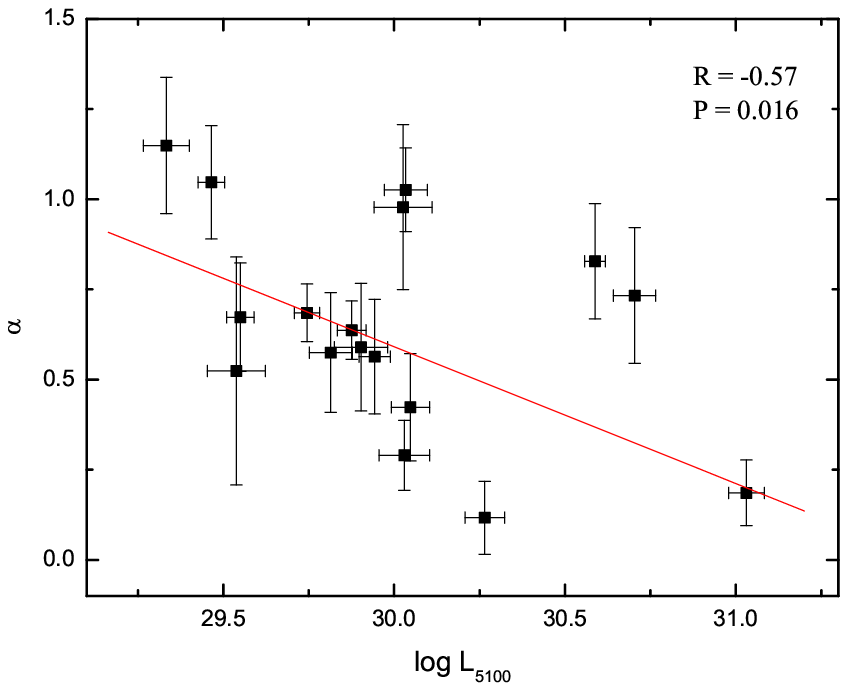}
   \includegraphics[width=80mm]{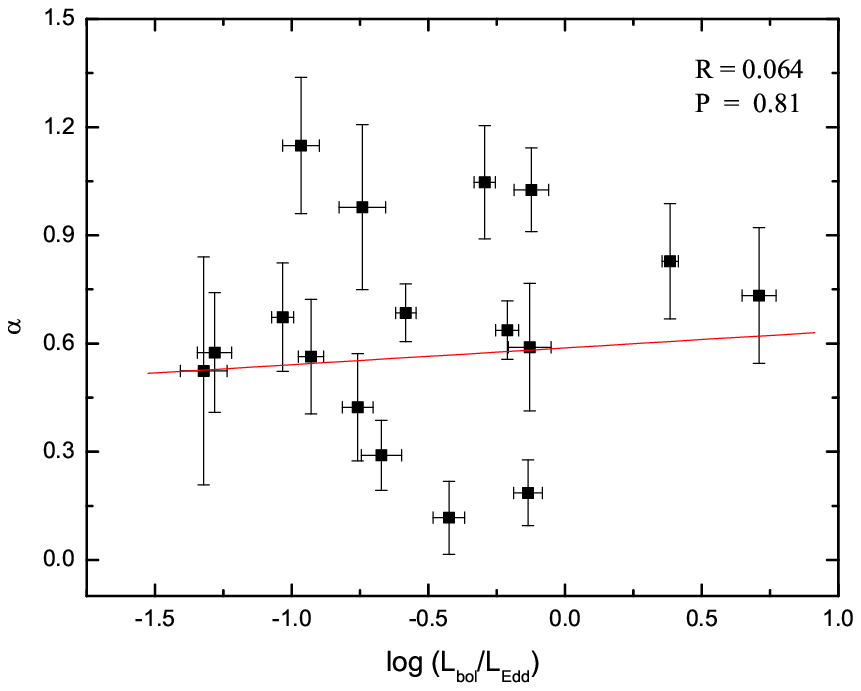}
   \caption{Left: the average spectral slope $\alpha$ versus the average
rest-frame 5100$\AA$ luminosity $\log L_ {5100}$ in units of
$ergs~s^{-1}Hz^{-1}$ for 17 PG QSOs. Right: the average spectral
slope $\alpha$ versus the average Eddington ratio for 17 PG QSOs.
The solid lines are our best fits. The correlation coefficient R
and the probability for rejecting the null hypothesis of no
correlation are showed in the right figure corners.}
   \label{}
   \end{figure*}

   \newpage

\begin{figure*}
   \centering
   \includegraphics[width=110mm]{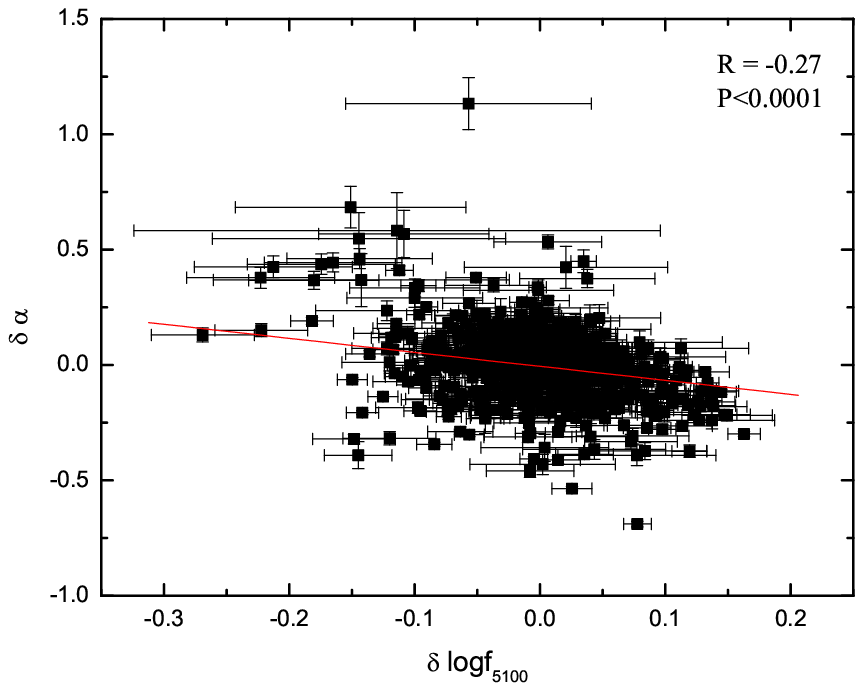}
   \caption{The spectral slope changes $\delta \alpha $ versus the
rest-frame 5100$\AA$ continuum flux variations $\delta \log f_
{5100}$. The solid line is our best fit.}
   \label{}
   \end{figure*}

\end{document}